\documentclass[aps,eqsecnum,groupedaddress,fleqn]{revtex4}
\usepackage{graphicx}
\usepackage{amssymb}
\newcommand{\be}{\begin{equation}}
\newcommand{\ee}{\end{equation}}
\newcommand{\bea}{\begin{eqnarray}}
\newcommand{\eea}{\end{eqnarray}}

\newcommand{\eqref}[1]{(\ref{#1})}
\newcommand{\F}{\phantom {1}}

\begin{document}



\title{A detailed study of the Abelian-projected SU(2) flux tube\\
and its dual Ginzburg-Landau analysis}

\author{Y. Koma}
\email[]{ykoma@mppmu.mpg.de}
\affiliation{Max-Planck-Institut f\"ur Physik, 
F\"ohringer Ring 6, D-80805 M\"unchen,  Germany}

\author{M. Koma}
\email[]{mkoma@mppmu.mpg.de}
\affiliation{Max-Planck-Institut f\"ur Physik, 
F\"ohringer Ring 6, D-80805 M\"unchen,  Germany}

\author{E.-M. Ilgenfritz}
\email[]{ilgenfri@physik.hu-berlin.de}
\affiliation{
Institut f\"ur Physik, Humboldt Universit\"at zu Berlin,
Newton-Str. 15, D-12489 Berlin,  Germany}

\author{T. Suzuki}
\email[]{suzuki@hep.s.kanazawa-u.ac.jp}
\affiliation{Institute for Theoretical Physics, Kanazawa University,\\
Kakuma-machi, Kanazawa, Ishikawa 920-1192, Japan}

\date{\today}

\begin{abstract}
The color-electric flux tube 
of Abelian-projected (AP) SU(2) lattice gauge theory 
in the maximally Abelian gauge (MAG) is revisited. 
It is shown that the lattice Gribov copy effect
in the MAG is crucial for the monopole-related 
parts of the flux-tube profiles.
Taking into account both the 
gauge fixing procedure and the effect of 
finite quark-antiquark distance properly,
the scaling property of the flux-tube profile is confirmed.
The quantitative relation between the 
measured AP flux tube and the flux-tube solution 
of the U(1) dual Abelian Higgs (DAH) model is also discussed.
The fitting of the AP flux tube in terms of the DAH flux tube
indicates that the vacuum can be classified as 
weakly type-I dual superconductor.
\end{abstract}

\maketitle

\baselineskip=0.55cm

\section{Introduction}

An intuitive explanation why a quark
cannot be isolated as a free particle
rests on the assumption that the QCD vacuum has the property
of a {\em dual}
superconductor~\cite{tHooft:1975pu,Mandelstam:1974pi}.
Analogously to electrically-charged Cooper pairs
being condensed in normal superconductors,
magnetically-charged monopoles
would be condensed in the QCD vacuum,
and the dual analogue of the Meissner effect
could be expected to occur.
In the result, for example, the color-electric flux
connecting a quark-antiquark ($q$-$\bar{q}$) system
would be squeezed into a quasi-one-dimensional 
flux tube~\cite{Abrikosov:1957sx,Nielsen:1973ve,Nambu:1974zg}.
This configuration provides
a linearly rising potential between
the quark and antiquark
such that a quark cannot be separated infinitely
from an antiquark  spending a finite amount of energy.

\par
Remarkably, lattice QCD simulations
with 't Hooft's Abelian projection~\cite{tHooft:1981ht}, typically
in the maximally Abelian gauge (MAG),
support this {\em picture} numerically.
The distributions of
the electric field and of the magnetic monopole currents,
which can be identified after Abelian projection,
have been measured in the presence of a static
$q$-$\bar{q}$ pair (represented by a Wilson loop) within
pure SU(2)~\cite{Singh:1993jj,Matsubara:1994nq} and
SU(3) lattice gauge theories~\cite{Matsubara:1994nq}.
It has been shown that the electric flux is confined in a
dual Abrikosov vortex due to
a monopole current circulating around the $q$-$\bar{q}$ axis,
signalling the dual Meissner effect.
More quantitatively,
the London penetration length of the electric field
has been studied systematically within
SU(2) lattice gauge theory~\cite{Cea:1995zt}.
These authors compared the penetration length in MAG with
a gauge invariant definition of the flux-tube profile.
They came to the conclusion that the penetration length
is gauge independent.
A large-scale simulation on a $32^{4}$ lattice at
a single value of $\beta=2.5115$ within 
SU(2) lattice gauge theory
has been performed next~\cite{Bali:1998gz},
applying the fine-tuned gauge
fixing algorithm (a mixture of overrelaxation (OR) algorithm and
a realization of {\em simulated annealing} (SA))~\cite{Bali:1996dm}
in order to fight the lattice Gribov copy problem in MAG and
applying a noise reduction 
technique, the smearing of spacelike link variables
before constructing a Wilson loop~\cite{Bali:1996dm}.
Through this study, the dual Amp\`ere law,
a relation between the monopole current and the
curl of electric field, has been confirmed with high accuracy.

\par
What is interesting and suggestive is that
these numerical results gave strong hints towards the existence
of a dual Abelian Higgs (DAH) model 
(we often call this the dual Ginzburg-Landau (DGL) model) 
as an effective model to deal with the QCD
vacuum~\cite{Suzuki:1988yq,Maedan:1988yi,Suganuma:1995ps}
and with quark-induced hadronic excitations of the vacuum.
In particular, the DAH model has a flux-tube solution
corresponding to a $q$-$\bar{q}$ system.
The mechanism of the flux-tube formation is
nothing but the dual Meissner effect.
The DAH model essentially contains three
parameters, the dual gauge coupling $\beta_{g}$, the
masses of the dual gauge boson $m_{B}$ and
of the monopoles $m_{\chi}$.
The inverses of these masses are identified as the London
penetration length and the Higgs coherence length, respectively.
These lengths determine the width of the flux tube.
The so-called Ginzburg-Landau parameter
is defined as the ratio between the two masses,
$\kappa = m_{\chi}/m_{B}$, where 
the vacuum for $\kappa < 1$~$(>1)$ is classified as
type-I (type-II) superconductor.

\par
Determining  the parameters of the DAH model,
based on the comparison between the
most elementary flux-tube profile measured
in non-Abelian lattice gauge theory
and the flux-tube solution of the DAH model,
is expected to be an important
source of information on the QCD-vacuum structure itself, which
would be helpful to learn about how the vacuum can
function as a dual superconductor.
Moreover, it might be useful in view of possible applications
of the model in more complex physical situations
(to describe baryons, for example in the SU(3) case).
The first motivation has been, more or less, common to the above
mentioned works and,  in fact,
such quantitative level of investigation has been attempted
in Refs.~\cite{Singh:1993jj,Matsubara:1994nq,Cea:1995zt,Bali:1998gz}.
For example, the lattice data have been fitted by
some approximate, 
analytical flux-tube solutions of the DAH model,
and in this way the GL parameter have been estimated.
The conclusions have varied, ranging from the vacuum 
belonging to the
borderline between a type-I and type-II superconductor,
with $\kappa \sim 1$, 
as claimed in Refs.~\cite{Singh:1993jj,Matsubara:1994nq},
to a classification of the vacuum as a type-I superconductor,
with $\kappa <1$, in Ref.~\cite{Bali:1998gz}.
The reanalysis of the
profile data in Ref.~\cite{Bali:1998gz} by fitting 
them with a numerical solution of the DAH model
has supported the case of $\kappa  \sim 1$~\cite{Gubarev:1999yp}.

\par
However, the systematic analysis of
lattice flux-tube data in MAG itself was incomplete such that
doubts have remained whether the resulting parameters
represent physical reality.
At least, one has to check several basic properties
of the lattice flux-tube profile before
one can seriously discuss the
implications of the extracted DAH parameters.
To mention the first, since one applies the MAG fixing,
the lattice Gribov copy effect should be controlled,
where the OR-SA algorithm would be 
helpful for this purpose~\cite{Bali:1996dm}.
At second, one should check the scaling property of the profile
with respect to changing the gauge coupling 
$\beta$ of the lattice simulation.
At third, one has to inquire how the profile behaves as a function
of the $q$-$\bar{q}$ distance.
One also should know how to compare the lattice flux-tube 
profile with the flux-tube solution in the DAH model,
if the above mentioned quantitative analysis is of interest.

\par
In our previous paper~\cite{Koma:2003gq},
which was mainly devoted to the duality relating
non-Abelian lattice theory on one hand with the 
DAH model on the other, we have
carefully studied the flux-tube structure
in the U(1) DAH model and
have confronted it with some related data from
our corresponding ongoing SU(2) lattice gauge
measurements in the MAG.
These studies, which will be discussed in the present paper
in much more detail, have been done using 
a $32^{4}$ lattice with the OR-SA algorithm and with 
the smearing technique as in Ref.~\cite{Bali:1996dm}.
Then, based on the Hodge decomposition of the Abelian
Wilson loop into the electric photon and the magnetic monopole parts,
we have found there that the 
Abelian-projected lattice flux tube consists of
two components, the Coulombic and the solenoidal electric field,
the latter being induced by the monopole currents circulating
around the $q$-$\bar{q}$ axis.
All this was in full analogy to
the structure of the DAH flux-tube solution.
We have also found~\cite{Koma:2003gq} 
that the Coulomb contribution cannot be neglected for any
flux-tube length practically accessible
in present-day lattice studies.

\par
In this paper we are going to present
all our results concerning the flux-tube profile within 
AP-SU(2) lattice gauge theory,
obtained in the MAG, in a more complete
way in order to meet the above requirements.
The strategy of our study has been the following one.
Measurements have been performed using a $32^{4}$ lattice
at various $\beta$ values ($\beta$=2.3, 2.4, 2.5115, and 2.6).
At first, we have investigated the lattice Gribov effect
by comparing the profiles
obtained from the widely used OR algorithm
and from the OR-SA algorithm.
In the context of the latter algorithm, 
we  have also investigated the
dependence on the number of gauge copies under investigation.
At second, we have studied the scaling property of the flux-tube
profile by comparing the profiles from
various $\beta$ values, keeping
the physical $q$-$\bar{q}$ distance approximately the same.
The physical scales, the lattice spacing $a(\beta)$
for different values of $\beta$, have been calculated through
the measurement of the corresponding non-Abelian 
string tension.
Throughout the profile measurements,
smearing has always applied 
to the spatial link variables 
before constructing a Wilson loop.
This procedure is 
meant to extract the profiles which effectively
belong to the ground state of a flux tube;
we have checked the (in)dependence of the flux-tube profile
on the temporal extension of the Wilson loop.
This effect has not received the due attention 
in the previous studies in
Refs.~\cite{Singh:1993jj,Matsubara:1994nq,Cea:1995zt,Bali:1998gz}.
This needs to be checked carefully when
the Wilson loop is used to represent a
static $q$-$\bar{q}$ source
and the ground state is of interest.
This procedure finally helps to reduce the noise.
In a final step, we have assessed the DAH parameters
by fitting the lattice data
against the numerical DAH flux-tube solution.
For this fit, we have not used the infinitely long
flux-tube solution as it has been done in previous
analyses~\cite{Singh:1993jj,Matsubara:1994nq,Cea:1995zt,
Bali:1998gz,Gubarev:1999yp}.
It should be noted that the use of  the infinitely long solution
would be suitable {\em only} for that part of the electric field
which is induced by the monopole part of the Wilson loop
with sufficiently large temporal length~\cite{Koma:2003gq}.
For our purpose to assess the DAH parameters, 
in this way
we have taken into full account the finite 
$q$-$\bar{q}$ length effect in the fit.

\par
The paper is organized as follows.
In section~\ref{sec:procedures}
we describe the procedures how to measure
the Abelian-projected SU(2) flux-tube profile in MAG.
Section~\ref{sec:numerical} presents the numerical results.
In section~\ref{sec:dah-fit} we describe the results of fitting
the lattice profiles by the DAH flux-tube solutions.
Section~\ref{sec:conclusion} 
is a summary and contains our conclusions.
Preliminary results of the studies 
summarized in the present paper
have been presented at the 
LATTICE2002 conference~\cite{Koma:2002uq}
and in Ref.~\cite{Koma:2003gq}.

\section{Numerical procedures}
\label{sec:procedures}

In this section, we explain how to measure the profile of a  
color-electric flux tube on the lattice within the maximally 
Abelian gauge (MAG).
We also develop the strategy to achieve a systematic, 
more detailed study of the flux-tube profile which 
takes into account the effect of the 
finite $q$-$\bar{q}$ length properly.
We restrict the explanations of the methods to the 
case of SU(2) gauge theory. 

\par
The numerical study of the flux-tube profile begins with 
the simulation of non-Abelian gauge fields.
A thermalized ensemble of SU(2) gauge configurations
$\{ U_{\mu} (m)\}$ is generated by simulating the 
standard  Wilson action
\bea
S_{\rm SU(2)}[U] =
\beta \sum_{m}\sum_{\mu <\nu}
\left \{ 1-\frac{1}{2} {\rm Re}
\; {\rm tr}[U_{\mu\nu}(m)] \right \}
\label{eq:wilson_action}
\eea
using the Monte Carlo heatbath method.
Here $U_{\mu\nu} (m) \in SU(2)$ are plaquette 
variables constructed in terms of link variables 
$U_{\mu} (m) \in SU(2)$ as
\bea
U_{\mu\nu} (m) \equiv U_{\mu} (m) U_{\nu}(m+\hat{\mu})
U_{\mu}^{\dagger} (m+\hat{\nu}) U_{\nu}^{\dagger}(m) \; .
\label{eq:plaquette}
\eea
The inverse coupling is given by 
$\beta = 4/e^{2}$.

\subsection{Maximally Abelian Gauge fixing}
\label{sec:MAG}

We put all the equilibrium configurations into the MAG.
One exploits the gauge freedom of the SU(2) link 
variables with respect to
gauge transformations $g(m)$ 
\bea
U_{\mu} (m) \mapsto U_{\mu}^{g} (m)
= g(m) U_{\mu} (m) g^{\dagger}(m+\hat{\mu}) 
\label{eq:gauge_trafo}
\eea
in order to achieve a maximum of the following gauge functional
\bea
R[U^g] = \frac{1}{8 V}\sum_{m,\mu} 
{\rm tr} \left \{  \tau_3 U_{\mu}^{g}(m) 
\tau_3 U^{g \; \dagger}_{\mu} (m)\right \} \; ,
\label{eq:gauge_functional}
\eea
where $V$ is the number of sites in the lattice.
The set of $g(m) \in SU(2)/U(1)$ for all site $m$ represents 
the MAG fixing gauge transformation 
defined on $m$.

\par
For the numerical task to find the ``optimal'' $g(m)$,
in the past mostly an overrelaxation (OR) 
algorithm has been used.
However, as it has been pointed out in the work 
in Ref.~\cite{Bali:1996dm},
the OR algorithm is prone to fall into the 
nearest  local maximum of Eq.~(\ref{eq:gauge_functional})
although the absolute maximum is of interest.
This is due to the existence of many local maxima,
which is  known as the lattice Gribov copy problem.
The only way known before to 
reduce the risk
of being trapped in a wrong maximum is to explore many 
such local maxima by repeating the OR algorithm, 
starting each time from a new random 
gauge copy of the
original Monte Carlo configuration;
the ensemble of $U_{\mu}^{g}(m)$ corresponding to only 
the highest of the 
achieved maxima was then understood as {\it the} gauge-fixed 
ensemble. 
It has then been proposed to use a ``simulated annealing (SA)''
algorithm with a  (final) OR algorithm
in order to prevent very poor maxima 
from entering the competition between 
gauge copies~\cite{Bali:1996dm}.
We may call this the OR-SA algorithm, 
which we mainly apply to fixing the MAG
in the present paper.
In the SA algorithm, the functional $R[U^{g}]$ is regarded as
a spin action
\bea
F(S)=R[U^{g}] =
\frac{1}{8 V}\sum_{m,\mu} 
{\rm tr} \left \{  S (m) U_{\mu}(m)
S(m+ \hat{\mu}) U^{\dagger}_{\mu} (m) \right \} ,
\eea
where
$S(m)=g^{\dagger}(m) \tau_{3} g(m)$
corresponds to spin variables.
The maximization of this functional is achieved by
considering the statistical system given by the
partition function
\bea
Z= \sum_{\{ S(m) \} } \exp \left [ \frac{1}{\tau} F(S) \right ]
\eea
with decreasing the auxiliary temperature $\tau \to 0$.
Practically, we first prepare a thermalized spin system at
a certain high temperature, which is 
decreased gradually according to some annealing schedule
until sufficiently low temperature is reached ($\tau \approx 0$).
Then, final maximization is done by 
means of the OR algorithms.
Notice that this algorithm only succeeds to escape from the
worst local maxima, such that the above-mentioned 
inspection of many gauge copies per Monte Carlo 
configuration cannot be avoided.

\par
After the MAG  fixing~\eqref{eq:gauge_trafo},
the SU(2) link variables 
$U_{\mu}^{g}(m)=U_{\mu}^{\mathit{MA}}(m)$ are factorized 
into a diagonal (Abelian) link variable 
$u_{\mu}(m) \in U(1)$ 
and the off-diagonal (charged matter field) parts 
$c_{\mu}(m)$, $c_{\mu}^*(m) \in SU(2)/U(1)$ 
as follows
\bea
U_{\mu}^{\mathit{MA}}(m)=
\left (
\begin{array}{cc}
\sqrt{1-|c_{\mu}(m)|^2} & -c_{\mu}^*(m)\\
c_{\mu}^*(m) & \sqrt{1-|c_{\mu}(m)|^2}
\end{array}
\right )
\left (
\begin{array}{cc}
u_{\mu}(m) & 0 \\
0 & u_{\mu}^* (m) 
\end{array}
\right ) \; .
\label{eqn:lat-Cartan-decompose}
\eea
The Abelian link variables  $u_{\mu}(m)$  are explicitly written as
\bea
u_{\mu} (m) =e^{i\theta_{\mu}(m)}
\quad  (\theta_{\mu} (m) \in [ -\pi, \pi ) ) \; .
\label{eq:Abelian_links}
\eea
The Abelian plaquette variables are constructed
from the phase $\theta_{\mu}(m)$ as
\bea
\theta_{\mu\nu}(m)
&\equiv& 
\theta_{\mu}(m) +\theta_{ \nu}(m+\hat{\mu})
- \theta_{\mu}(m+\hat{\nu}) -\theta_{\nu}(m)
 \quad ( \theta_{\mu\nu}(m) \in [-4\pi, 4 \pi) ) \; ,
\label{eq:Abelian_plaquette}
\eea
which  can be decomposed into a regular part
$\bar{\theta}_{\mu\nu}(m) \in [-\pi,\pi)$
and a singular (magnetic Dirac string) 
part $n_{\mu\nu} (m)= 0,\pm 1,\pm 2$ 
as follows
\bea
\theta_{\mu\nu}(m)
&\equiv& 
\bar{\theta}_{\mu\nu}(m) +2\pi n_{\mu\nu}(m) \; .
\label{eq:Abelian_plaquette_decomposition}
\eea
The field strength is defined by 
$\bar{\theta}_{\mu\nu} (m)
=\theta_{\mu\nu}(m) -2 \pi n_{\mu\nu}(m)$.
Following DeGrand and Touissaint~\cite{DeGrand:1980eq}, 
magnetic monopoles are extracted 
from the magnetic Dirac string sheets as their boundaries
\bea
k_{\mu}(\tilde{m})
=
- \frac{1}{2} \varepsilon_{\mu\nu\rho\sigma}
\partial_{\nu} n_{\rho\sigma} (m+\hat{\mu})
\qquad (  \varepsilon_{1234}=1 ) \; ,
\label{eq:magnetic_current_definition}
\eea
where $k_{\mu} (\tilde{m})= 0,\pm 1,\pm 2$ and 
$\tilde{m}$ denotes the dual site defined by
$\tilde{m} = m+ (\hat{1} + \hat{2} + \hat{3} + \hat{4})/2$.
Note that the monopole current satisfies a conservation law 
$\partial_{\mu}' k_{\mu}(\tilde{m}) = 0$ 
formulated in terms of the backward derivative $\partial_{\mu}'$.

\subsection{Correlation functions of Wilson loops
involving local probes}
\label{sec:correlators}

To find the flux-tube profile, one needs to measure the 
expectation value of a local probe ${\cal O}(m)$ with an external 
source as $\langle {\cal O}(m)\rangle_j$ (in our case $j$ corresponds to an
Abelian Wilson loop). 
Based on the path integral representation of 
$\langle {\cal O}(m)\rangle_j$, it can
be rewritten as the ratio of $\langle W_{A}{\cal O}(m)\rangle_0$ and 
$\langle W_{A}\rangle_0$, where the subscript $0$ means the 
expectation value in the vacuum without such a 
source~\cite{Koma:2003gq}:
\bea
\langle {\cal O}(m)  \rangle_{j}
= 
\frac{\langle   W_{A} {\cal O}(m) \rangle_{0} }
{\langle  W_{A} \rangle_{0} } \; .
\label{eq:Wilson_loop_correlators}
\eea
The Abelian Wilson loop is defined in terms of the $u_{\mu}(m)$ 
\bea
W_{A} (L) = \prod_{m \in L} u_{\mu}(m)
=\exp [i \! \sum_{m \in L} \theta_{\mu}(m) ] \; .
\label{eq:Wilson_loop_definition}
\eea
The local operators ${\cal O}(m)$ that we need to describe the 
structure of the flux tube are 
an electric field  operator
\bea
i \bar{\theta}_{i4}(m)
= i(\theta_{i4}(m)  -2 \pi n_{i4} (m)) , 
\label{eq:electric_field}
\eea
and a monopole current operator
\bea
2 \pi  i k_{i} (\tilde{m}) \; ,
\label{eq:monopole_current}
\eea
where only spatial directions $i=1,2,3$ are of interest.
To avoid the contamination from higher states as much as possible, 
we have inserted the local probe ${\cal O}(m)$ 
at $t=t_{m}=T/2$ to minimize the
effect from the boundary of the Wilson loop at $t=0$ and $T$.
The local field operators are 
then evaluated over the whole 
$x$-$y$ midplane erected in the center
of the spatial extension of the Abelian Wilson loop ($z=z_{m}=R/2$).
In other words,
the coordinates of the local operator are $m=(x,y,z_m,t_m)$ running over 
the midplane of the flux tube between a quark and an antiquark.

\subsection{Decomposition of the Abelian Wilson loop}
\label{sec:decomposition}

In order to see the composed
structure of the flux-tube profiles,
it is useful to  apply the Hodge decomposition to 
the Abelian Wilson loop, which allows us to define
the photon and monopole Wilson loops.
We have shown in the previous work~\cite{Koma:2003gq}
for the electric field profile 
that the photon Wilson loop induces 
exclusively the Coulombic electric field 
while the monopole Wilson loop creates 
the solenoidal electric field.
At the same time, 
concerning the monopole current profile,
the photon Wilson loop 
is not correlated with the monopole currents, while exclusively the
monopole Wilson loop is responsible for 
the monopole current signal.

\par
We explain the decomposition using lattice 
differential form notations~\cite{Chernodub:1997ay}.
The Abelian Wilson loop in Eq.~\eqref{eq:Wilson_loop_definition}
is written as $W_{A}=\exp [i(\theta,j)]$, where $\theta (C_{1})$
and $j(C_{1})$ are the Abelian link variables and
the closed electric current.
The Hodge decomposition of Abelian link variables leads to
\bea
\theta 
&=& \Delta^{-1} \Delta \theta 
= \Delta^{-1} (d\delta + \delta d) \theta \nonumber\\*
&=&
\Delta^{-1} d\delta \theta +
\Delta^{-1}\delta \bar{\theta} + 2 \pi \Delta^{-1}\delta n 
 \; ,
\label{eq:first}
\eea
where the second and third terms are identified as
the photon link ($\theta^{ph}= \Delta^{-1} \delta \bar{\theta}$)
and monopole link ($\theta^{mo}=2 \pi \Delta^{-1}\delta n$) 
variables, respectively.
We do not need to fix the Abelian gauge in order to specify the first
term in the second line, since it
does not contribute to the Abelian Wilson loop due to $\delta j=0$:
$(\Delta^{-1} d \delta \theta,j)
=  (d \Delta^{-1} \delta \theta,j)
=  (\Delta^{-1} \delta \theta, \delta j)=0$.
Note that 
$\Delta$ is a lattice Laplacian and 
$\Delta^{-1}$ is its inverse, corresponding to the 
lattice Coulomb propagator.
We have used $d \theta =  \bar{\theta} + 2 \pi n $
(see, Eq.~\eqref{eq:Abelian_plaquette_decomposition})
at the last equality.
Inserting Eq.~\eqref{eq:first} into the expression of $W_{A}$,
the Abelian Wilson loop is decomposed as 
\bea
W_{A}[j] = \exp [i(\theta^{ph},j)] \cdot \exp [i(\theta^{mo},j)] 
\equiv 
W_{\mathit{Ph}}[j]  \cdot W_{\mathit{Mo}}[j] \; .
\label{eq:Wilson_loop_decomposition}
\eea
We call $W_{\mathit{Ph}}$ and  $W_{\mathit{Mo}}$ 
the photon Wilson loop and the monopole Wilson loop.
They are separately used to evaluate the photon and monopole parts
of the profiles.

\subsection{Smearing of spatial links}
\label{subsec:smearing}

The shape of the flux-tube profile induced by a Wilson loop depends on 
its size, $R \times T$, where $R$
corresponds to the $q$-$\bar{q}$ distance
and $T$ the temporal extension.
This means that the profile 
is influenced not only by the ground state but also
by excited states, when a Wilson loop is used as
an external source.
If one is interested in the ground state, 
{\it e.g.}, for the comparison with the
flux-tube solution in the three-dimensional 
DAH model,
in principle, one needs to know the profile at
$T \to \infty$.
However, it is practically 
impossible to take this limit 
due to the finite lattice volume.

\par
Smearing is a useful technique
to extract the profiles which belong to the 
ground state of a flux tube 
effectively even with  a finite $T$.
We then see remarkable noise reduction when 
the size of the Wilson loop is large.
The procedure is as follows.
Regarding the fourth direction as the Euclidean 
time direction, we perform the following step
several ($N_{s}$) times
for the spacelike Abelian link variables:
\bea
 \exp [i \theta_{i}(m) ]
 \mapsto
 \alpha \exp[ i \theta_{i}(m) ]
+ \sum_{j \ne i}
\exp  \left [ i (\theta_{j}(m) +\theta_{i}(m+ \hat{j})
- \theta_{j}(m+\hat{i})  ) \right ]   \; ,
\label{eq:smearing_step}
\eea
where $i,j = 1,2,3$ and 
$\alpha$ is an appropriate smearing weight.

\par
To find an appropriate set of parameters ($\alpha$, $N_{s}$),
one needs to investigate the $T$-dependence of 
several quantities like
the ground state overlap and the $q$-$\bar{q}$ potential.
The emerging shape of the profile also should be checked
for the effect of smearing.
A numerical example at $\beta=2.5115$ is shown in 
Appendix~\ref{sec:smearing-detail}.
We notice that this procedure seems to have 
practical limitations 
which become visible in the flux-tube profile measurement.
For the profile extracted with Wilson loops of size $R \leq T$,
it works very well
with large class of the parameter set ($\alpha$, $N_{s}$).
We could easily observe $T$-independence of the 
profile  within the numerical error.
On the other hand, for the 
Wilson loops of size $R > T$, 
an extremely fine-tuned parameter set is required
for smearing. 
However, we did not spend full effort to fix it.

\section{Numerical results}
\label{sec:numerical}

In this section, we present numerical results
of the flux-tube profiles measured 
over the $x$-$y$ midplane of the $q$-$\bar{q}$ system 
(separated in $z$ direction) using
the Abelian,  photon and monopole Wilson loops.
We are going to clarify 
{\it i}) the lattice Gribov copy effect associated with
the MAG fixing procedure and
{\it ii}) the scaling property.

\par
The numerical simulations were done at
$\beta=$ 2.3, 2.4, 2.5115, and $\beta=2.6$.
The lattice volume was always $V=32^{4}$.
We have used 100 configurations for measurements.
We have stored them after 3000 thermalization sweeps,
and they were separated by 500 Monte Carlo updates. 
To study ({\it i}), 
we have generated several numbers of gauge copies
from a given SU(2) configuration 
by random gauge transformations,  
each of which has undergone the OR-SA algorithm in the process of MAG fixing.
The SA algorithm itself 
is applied with the temperature 
decreasing from $\tau =2.5$ to $\tau =0.01$.
After that the OR algorithm is adopted with a 
certain convergence criterion.
As the number of gauge copies,
we have chosen $N_{g}=$ 5,~10~and~20,
and have stored the configurations which  
provide the best maximal value of the
gauge functional Eq.~\eqref{eq:gauge_functional} 
within these~$N_{g}$.
We have also stored the same number of 
configurations (=100) from the OR algorithm in the MAG fixing
with the same stopping criterion 
as in the OR-SA case, where 
always the first copy has been accepted ($N_{g}=1$).
To study ({\it ii}), we have used the 
configurations from
the sample based on~$N_{g}=$ 20 copies.
The Abelian smearing parameters 
have been chosen  $N_{s}=8$ for $\alpha=2.0$.
With this choice,  the temporal length independence
of the profiles induced by the 
Abelian Wilson loop  is achieved 
within errors, at least for $R \leq T$.
The same procedure was also applied to the spacelike
photon and monopole link variables before
constructing each type of Wilson loop.

\subsection{Fixing the physical scale and 
choosing the flux-tube lengths}

The physical reference scale, 
the lattice spacing $a(\beta)$,
has been determined from the non-Abelian string 
tensions $\sigma_L$ by fixing $\sqrt{\sigma_{phys}} 
= \sqrt{\sigma_L}/a \equiv 440$ MeV.
The non-Abelian string tension has been evaluated 
by measuring expectation values of non-Abelian Wilson
loops with optimized non-Abelian smearing.
The emerging potential 
has been fitted to match the form 
$V(R) = C - A/R + \sigma_L\; R$.
The resulting (dimensionless) lattice string tensions and the 
corresponding lattice spacings $a(\beta)$ in 
units of fm are shown in Table.~\ref{tab:scale}.

\begin{table}[t]
 \centering
\caption{The non-Abelian string tensions and 
corresponding lattice spacings $a(\beta)$
for $\beta =$ 2.3, 2.4, 2.5115, and 2.6 estimated by
the relation $a(\beta)=\sqrt{\sigma_{L}/\sigma_{phys}}$
with $\sqrt{\sigma_{phys}}=440$ MeV.}
\begin{tabular}{|l|l|l|}
    \hline
    $\beta$ & $\sigma_L$ & $a(\beta)$ [fm]  \\
    \hline
    2.3 & 0.144(3)\F & 0.170(2)   \\
    \hline
    2.4 & 0.0712(5) & 0.1197(4)   \\
    \hline
    2.5115 & 0.0323(4) & 0.0806(5)   \\
    \hline
    2.6 &  0.0186(2)  &  0.0612(5) \\
      \hline
\end{tabular}
\label{tab:scale}
\end{table}

\begin{figure}[t]
\centering
\includegraphics[width=10cm]{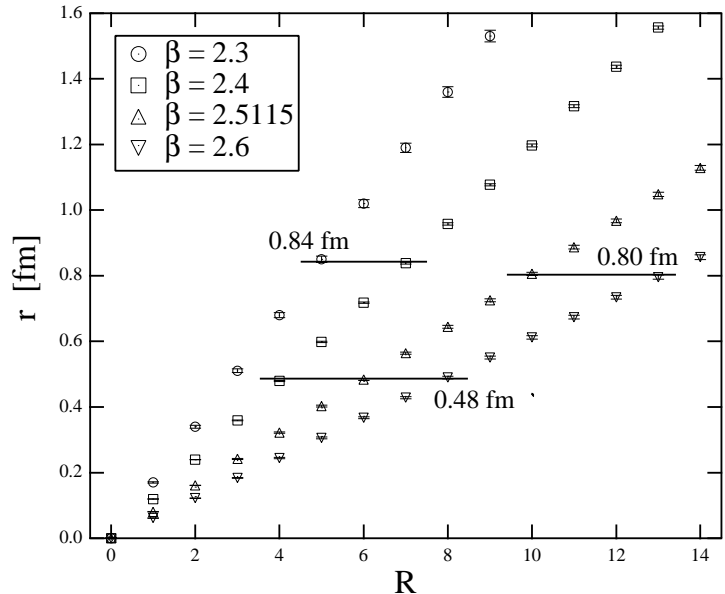}
\caption{Estimate of the physical $q$-$\bar{q}$ distance
$r$ in units of fm as a function of lattice distance $R$ for
various $\beta =$ 2.3, 2.4, 2.5115, and 2.6 used in this study.}
\label{fig:qqbar_length}
\end{figure}

\par
To compare the profiles from various $\beta$
values, it is quite important to put data into groups close to
almost the same physical $q$-$\bar{q}$ distance 
because the finite length effect of the flux-tube system 
has to be studied simultaneously~\cite{Koma:2003gq}.
One might naively expect that the flux-tube profile has a good 
translational-invariant property along the 
$q$-$\bar{q}$ axis so that 
the difference in length does not matter when one 
follows the change in lattice scale.
However, as shown in our previous 
work~\cite{Koma:2003gq}, the finite length effect is not negligible
as long as the photon part of the profile still 
contributes to the total profile. 
In Fig.~\ref{fig:qqbar_length} we then plot, 
for the four $\beta$-values at our disposal, 
the physical length $r = R a(\beta)$ in units of fm for various
choices of the integer lattice flux-tube 
length $R$.
This information is taken into account when we
study the scaling property of the flux-tube profile.

\subsection{Assessment of the lattice Gribov problem}

We investigate how the flux-tube profile depends on the 
lattice Gribov copy effect due to the MAG fixing.
As shown in Ref.~\cite{Bali:1996dm}, the density of 
monopole currents (in vacuum) is sensitive to the gauge fixing procedure; 
it decreases when larger $R[U^g]$ is achieved.
Therefore, we expect that 
the monopole-related part of the profile 
crucially depends on the quality of the gauge fixing procedure.

\par
\begin{figure}[!t]
\includegraphics[width=14cm]{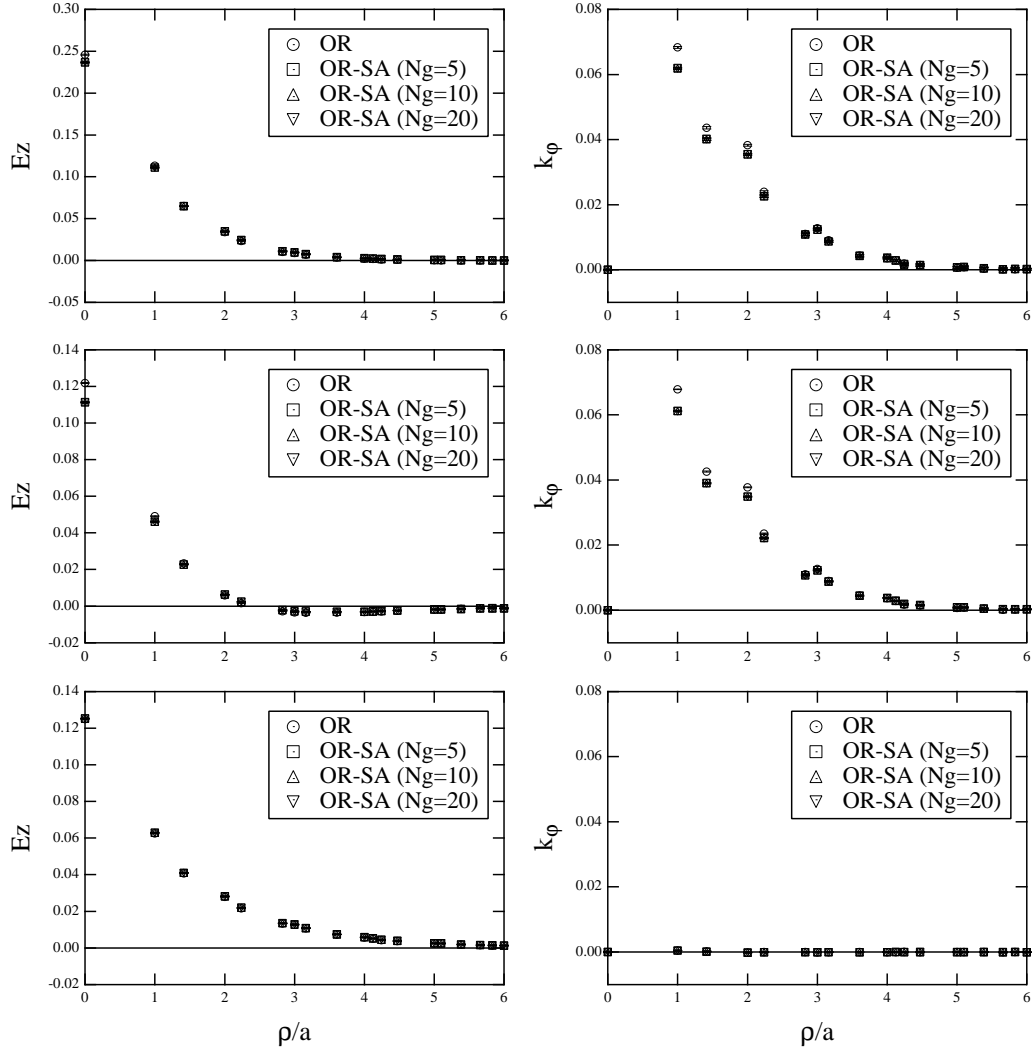}
\caption{The profiles of electric field (left) and monopole current (right) 
at $\beta=2.3$ for the $W(3,3)$ quark-antiquark system 
for gauge fixing according to OR and OR-SA algorithms,
respectively, and the dependence on the number of gauge copies in the OR-SA case.
Non-integer radiuses appear due to off-axis distances 
from the flux-tube axis.  
Upper, middle, and lower figures 
refer to correlations with the full Abelian, 
the monopole and the photon Wilson loops.}
\label{fig:gdep_b23_sh}
\end{figure}

In Fig.~\ref{fig:gdep_b23_sh} we show the electric 
field and monopole current profiles as observed 
over the flux-tube midplane at $\beta=2.3$ for $W(R,T)=W(3,3)$.
Profiles, both with the use of 
the OR and the OR-SA algorithms, are presented.
The dependence on the number of gauge copies under exploration 
is also investigated
for the case of the latter algorithm.
The upper, middle, and lower  
figures are the profiles from the Abelian, the monopole 
and the photon Wilson loops, respectively.

\par
We observe that the electric field and
the monopole current profiles
(except from the photon Wilson loop) 
are overestimated if the OR algorithm is applied. 
A possible explanation of this behavior 
is the following.
The correlation between the monopole Wilson loop and
the monopole currents are enhanced artificially
due to denser monopole current system owing to
the imperfect gauge fixing, which  
results in a larger contribution to the monopole current profile.
Then, the strongly circulating monopole 
current around the $q$-$\bar{q}$ 
axis induces a strong solenoidal electric field.
In this way, the electric field profile is also
overestimated when the OR algorithm is adopted.
It is interesting to note that 
since the photon Wilson loop 
is not correlated with the monopole currents, 
the corresponding electric field profile is insensitive
to the  Gribov copy problem.
Finally, the impact of the Gribov copy problem 
on the monopole part is inherited 
also by the total flux-tube profile 
measured by the Abelian Wilson loop.
Notice that the number of gauge copies
to which the OR-SA algorithm is applied does not 
drastically change the profiles compared with
the change from OR to OR-SA algorithm.
This suggests that the 
tentative maxima successfully 
anticipated at the end of SA algorithm 
do not strongly differ in the monopole density.  

\par
\begin{figure}[t]
\includegraphics[width=14cm]{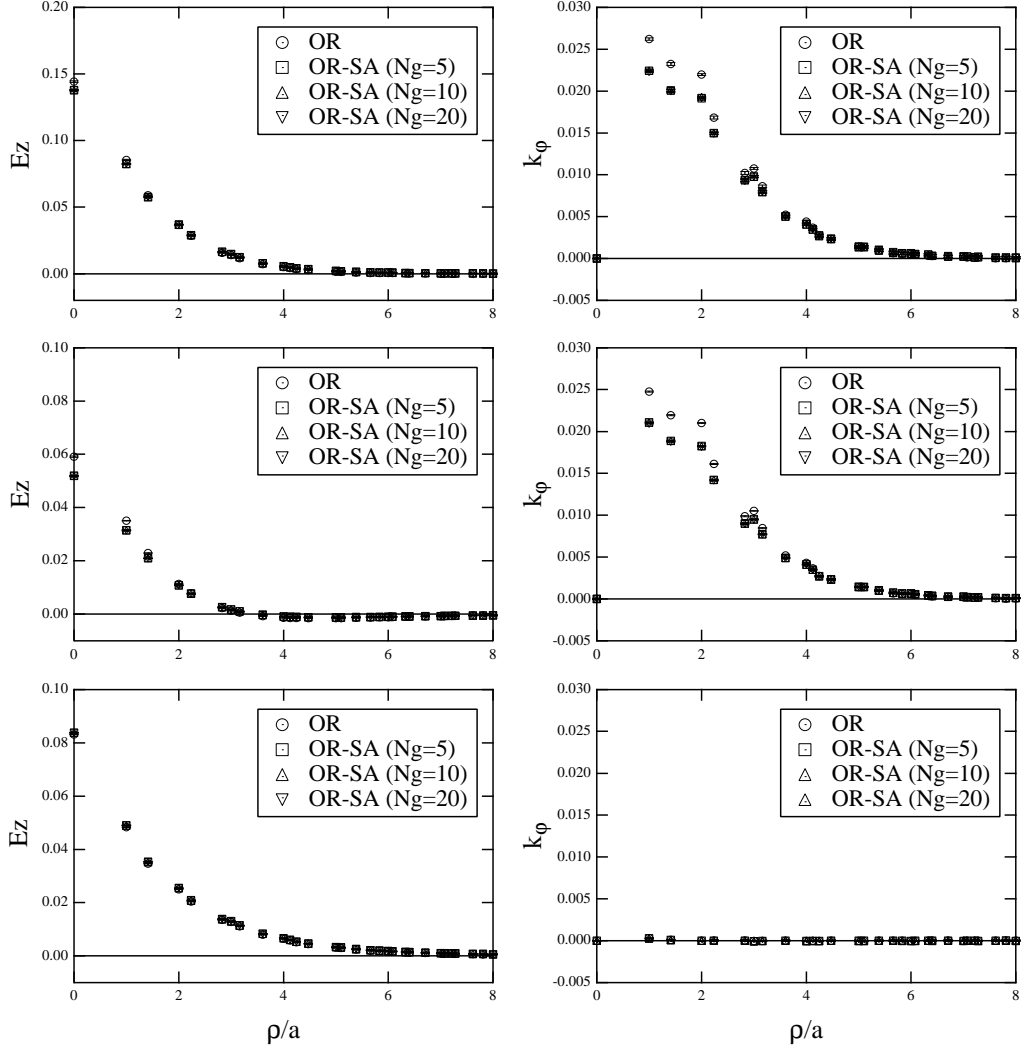}
\caption{The same plot as in Fig.~\ref{fig:gdep_b23_sh}
at $\beta=2.4$ for $W(4,4)$.}
\label{fig:gdep_b24_sh}
\end{figure}
\begin{figure}[t]
\includegraphics[width=14cm]{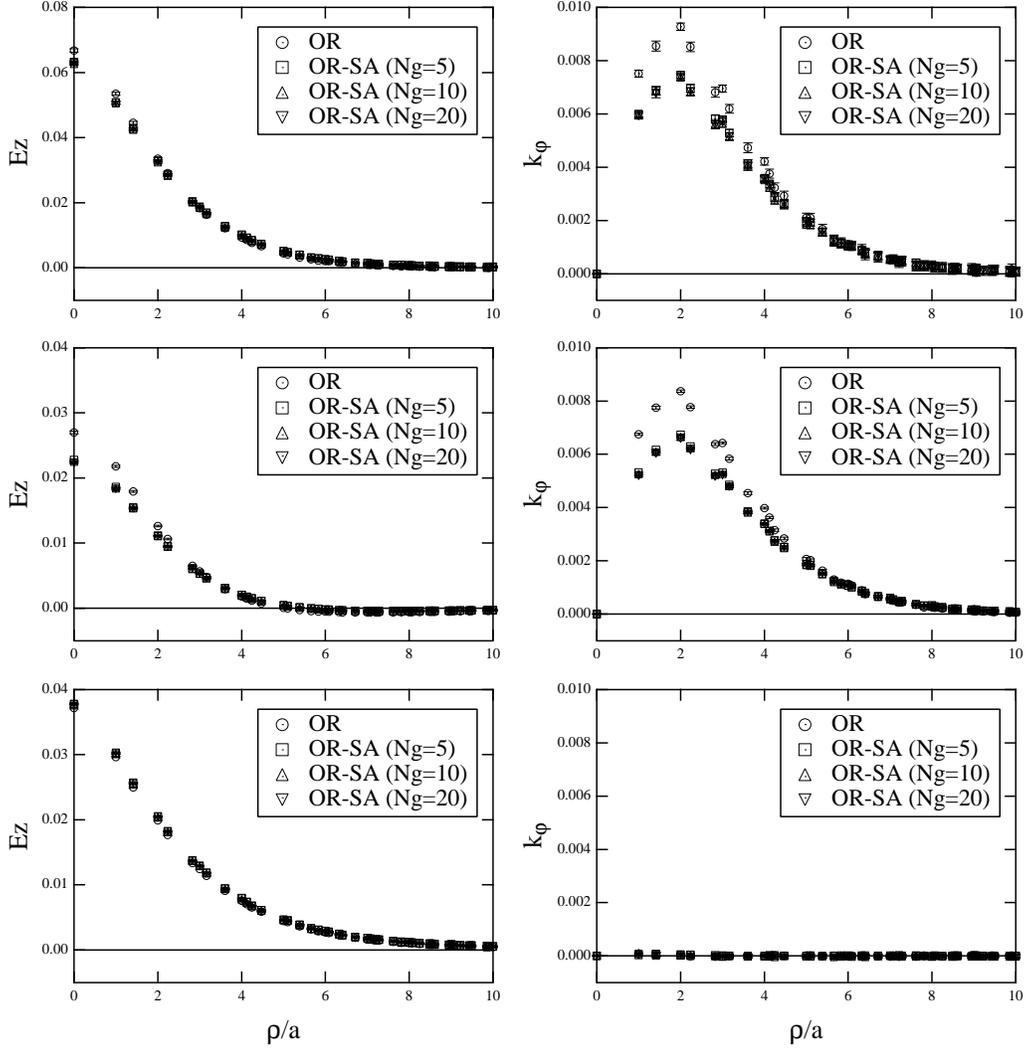}
\caption{The same plot as in Fig.~\ref{fig:gdep_b23_sh}
at $\beta=2.5115$ for $W(6,6)$.}
\label{fig:gdep_b251_sh}
\end{figure}
\begin{figure}[t]
\includegraphics[width=14cm]{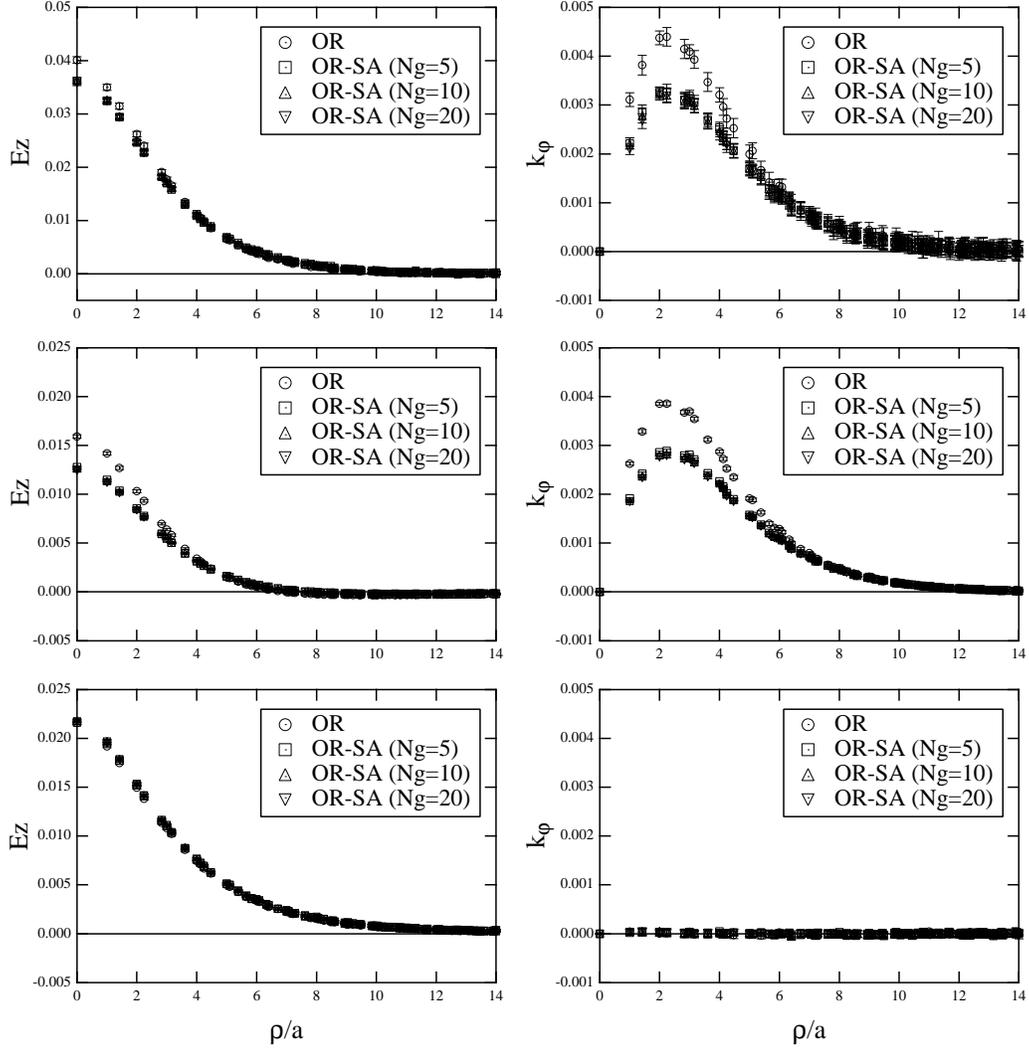}
\caption{The same plot as in Fig.~\ref{fig:gdep_b23_sh}
at $\beta=2.6$ for $W(8,8)$.}
\label{fig:gdep_b26_sh}
\end{figure}

In Figs.~\ref{fig:gdep_b24_sh},~\ref{fig:gdep_b251_sh} 
and~\ref{fig:gdep_b26_sh} we show the same plot as in 
Fig.~\ref{fig:gdep_b23_sh} for 
other $\beta$ values, correspondingly choosing the 
Wilson loops: 
at $\beta=2.4$ for $W(4,4)$, at $\beta=2.5115$ for 
$W(6,6)$  and at $\beta=2.6$ for $W(8,8)$.
Here, the physical sizes of 
the respective Wilson loops are
approximately the same (0.48 fm $\times$ 0.48 fm).
We find that with increasing $\beta$ (approaching to the 
continuum limit), the difference between the profiles from
the OR algorithm and the OR-SA algorithm
becomes clearer, {\it i.e.}
the effect of the Gribov copy problem becomes 
more significant.
Since the continuum limit is of interest,
one needs to take care of this problem as 
already emphasized in Ref.~\cite{Bornyakov:2001ux}.

\clearpage
\subsection{Does the flux-tube profile satisfy scaling ?}

\begin{figure}[t]
\includegraphics[width=14cm]{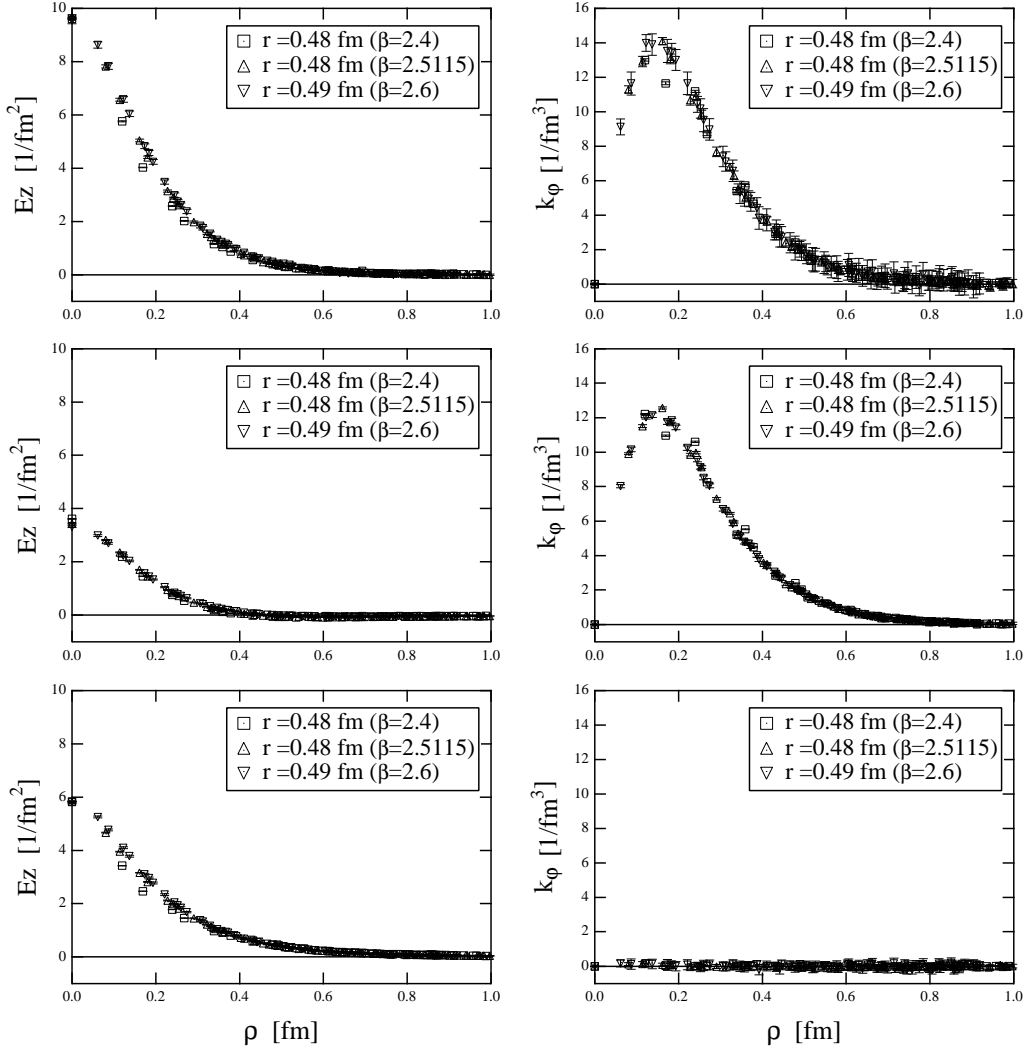}
\caption{The $\beta$ dependence of 
the profiles of the electric field (left) and the monopole current (right) 
for $q$-$\bar{q}$ distance 
$r \sim $ 0.48 fm as a function of the flux-tube 
radius $\rho$ given in units of fm 
(see Table~\ref{tab:scale} and Fig.~\ref{fig:qqbar_length}).}
\label{fig:scale_short}
\end{figure}

\begin{figure}[t]
\includegraphics[width=14cm]{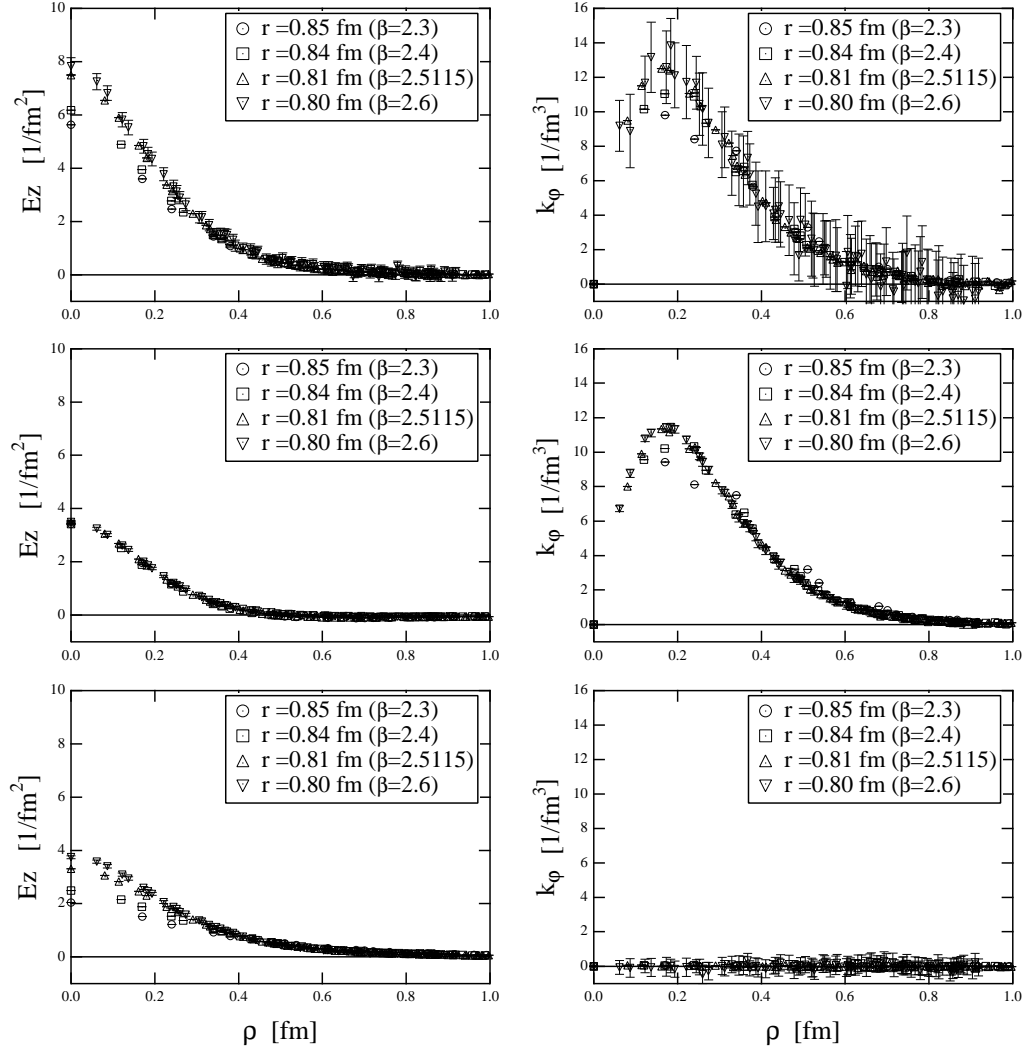}
\caption{The same plot as in Fig.~\ref{fig:scale_short}
for $q$-$\bar{q}$ distance $r \sim $ 0.80 and 0.84 fm 
(see Table~\ref{tab:scale} and Fig.~\ref{fig:qqbar_length}).}
\label{fig:scale_middle}
\end{figure}

We investigate the scaling property for groups
of $q$-$\bar{q}$ distances according to 
Fig.~\ref{fig:qqbar_length} using the 
best MAG-fixed configurations with $N_{g}=20$.
We choose three sets of physical distances:
\begin{itemize}
\item at $r \sim 0.48$ fm: 
from $\beta=$2.4 with $W(R,T)=W(4,4)$, 
$\beta=$2.5115 with $W(6,6)$,
and $\beta=$2.6 with $W(8,8)$,
\item at $r \sim 0.80$ fm:
from $\beta=$2.5115 with $W(10,6)$
and $\beta=$2.6 with $W(13,8)$.
\item   at $r \sim  0.84$ fm: 
from  $\beta=$2.3 with $W(5,3)$
and $\beta=$2.4 with $W(7,4)$.
\end{itemize}
Here, the physical size of the 
temporal extension of the Wilson loop 
is taken  approximately the same among all sets.
This choice is made to normalize the systematic uncertainty
for the flux-tube profile which might come from 
finite $T$ even after smearing,
especially for Wilson loops having $R > T$,
which reflect contributions from excited states.
We did not attempt to include profiles from such Wilson 
loops into the fit in Sec.~\ref{sec:dah-fit}.
The first set is shown in Fig.~\ref{fig:scale_short}.
The other two sets are plotted together in 
Fig.~\ref{fig:scale_middle}.

\par
We find that both the electric and monopole current 
profiles measured at different $\beta$ values from 
the interval 2.3 to 2.6 scale properly for each 
of the three groups of $q$-$\bar{q}$ distances.
The remaining minor differences can
be blamed to small differences in $q$-$\bar{q}$
distance and uncontrolled smearing effects.
We also observe the following properties.
Although the rotational invariance around 
$q$-$\bar{q}$ axis is poor for small $\beta$,
it is recovered with increasing $\beta$.
The electric field profile from the photon 
Wilson loop is very sensitive to the change 
of the $q$-$\bar{q}$ distance.
Clearly, the shape of the electric field profiles from the photon 
Wilson loop in Figs.~\ref{fig:scale_short}
and~\ref{fig:scale_middle} are different; 
this part of the electric field drastically decreases 
with increasing $q$-$\bar{q}$ distance.
On the other hand, the electric field profile from
the monopole Wilson loop 
remains almost the same with increasing $q$-$\bar{q}$
distance.
The difference of the electric field profile
coming from the full Abelian Wilson loop for different $q$-$\bar{q}$ 
distance can be explained by the change of the 
photon contribution.
The large error of the monopole current profile
from the Abelian Wilson loop in Fig.~\ref{fig:scale_middle} 
is due to the large size of the Wilson loop, $13\times 8$, at $\beta=$ 2.6.
The statistics is not sufficient in this case.
However, it is interesting to find that the decomposition 
of the Abelian Wilson loop into the photon and monopole parts 
helps to see a clear signal even with 
a number of configurations 
which is normally used for smaller Wilson loops.

\clearpage

\clearpage

\clearpage
\section{Fitting with the U(1) DAH flux tube}
\label{sec:dah-fit}

In this section, 
we discuss the quantitative relation between the
extracted AP-SU(2) flux tube and the classical flux tube 
of the dual Abelian Higgs model (the DAH flux tube)
through a $\chi^{2}$ fit of the former profile by the latter.
In the fit, we take into account both 
the electric field and the monopole current 
profiles simultaneously.

\subsection{The dual lattice formulation of the DAH model}

The DAH flux-tube profile is calculated
within the {\em dual lattice formulation} of the
{\it three dimensional DAH model} in order to 
mimic eventual lattice discretization 
effects in the fit~\cite{Koma:2000hw}.
The lattice DAH action is
\bea
S_{\rm DAH} =
\beta_{g}
\sum_{m} 
\left [ 
\frac{1}{2} \sum_{i < j} F_{ij}(m)^{2}
 + 
\frac{m_B^2}{2} \sum_{i=1}^{3}
\left | \Phi (m) -e^{i B_{i}(m)}  \Phi (m+\hat{i}) \right |^2
 +
\frac{m_B^2 m_{\chi}^2}{8}\left ( |\Phi (m)|^2 - 1 \right )^2
\right ],
\eea
where $F_{ij}$ is the dual field strength
\bea
F_{ij}(m)
=
B_{i}(m) + B_{j} (m+\hat{i})-B_{i} (m+\hat{j} ) -  B_{j}(m)
-
2 \pi \Sigma_{ij}(m) \, .
\eea
$B_{i}(m)$ and $\Phi(m)$ are the dual gauge field and 
the complex-valued scalar monopole field.
The electric Dirac string $\Sigma_{ij}$
in the dual field strength 
reflects the actual length of the flux tube.
For instance, for a straight flux tube along the $z$ direction,
$\Sigma_{12}=1$ for all plaquettes penetrated by this flux tube, 
otherwise $\Sigma_{ij}=0$~\cite{Koma:2000hw}.
This action contains three parameters: 
the dual gauge coupling $\beta_{g} = 1/g^{2}$,
the dual gauge boson mass $m_{B}= \sqrt{2}g v$,
and the monopole mass $m_{\chi}= 2 \sqrt{\lambda} v$.
Here $v$ corresponds to the monopole condensate,
and $\lambda$ is the self-coupling of the monopole field.
Writing the masses in terms of $g$, $v$ and $\lambda$ 
is more familiar in the continuum form of the DAH model.
Note that in the lattice formulation, all fields and 
parameters are  dimensionless. 
The type of dual superconductivity is
characterized by the Ginzburg-Landau parameter,
$\kappa =m_{\chi}/m_{B}$.
In this definition, 
the cases $\kappa  <1$~$(>1)$ are
classified as type I (type II) vacuum.

\par
The flux-tube solution is obtained 
by solving the field equations.
The equation for  $B_{i}(m)$ is
given by $\frac{\partial S_{\rm DAH}}{\partial B_{i}(m) }= 
\beta_{g}X_{i}(m)=0$.
Similarly, the equations for the monopole field are
$\frac{\partial S_{\rm DAH}}{\partial \Phi^R (m)}=
\beta_{g} m_B^2  X^R (m)=0$
and $\frac{\partial S_{\rm DAH}}{\partial \Phi^I (m)}=
\beta_{g} m_B^2  X^I (m)=0$.
The superscripts $R$ and $I$ refer to
the real and imaginary parts of the complex 
scalar monopole field.
The explicit form of the field 
equations $X_{i}(m)$, $X^R (m)$, and  $X^I (m)$
are given in Appendix~\ref{sec:fieldeqs}.
To solve the field equations numerically,
we adopt a relaxation algorithm {\it a la} Newton and Raphson 
by taking into account
the second derivative of the action with 
respect to each field~\cite{Koma:2000hw}.
We iterate this procedure until the 
conditions
$\sum_{m} \sum_{i=1}^{3} ( X_{i}(m) )^{2} < 0.0001$ 
and $\sum_{m} \{ ( X^R (m) )^{2}+ ( X^I (m) )^{2} \}  < 0.0001$
are satisfied
and the change of the action for one iteration
step $\Delta S_{\rm DAH} < 0.001$.
Within the possible range of the DAH parameters, we find 
that the solution is well-converged.

\par
The strategy of the fit is as follows.
We fit the flux-tube profile induced 
by the Abelian Wilson loop with the DAH flux tube.
Since we want to use the $T$-independent profile and it can only be
$T$-independent if $R \leq T$, we are restricted for this purpose (profile)
to $R \leq  T$.
The electric field and monopole current 
profiles of the DAH flux tube is
$\sqrt{\beta_{g}}\varepsilon_{ijk}  F_{jk}$ and 
$\sqrt{\beta_{g}} K_{i}$ (see, Appendix~\ref{sec:fieldeqs}),
which are regarded as representing $\bar{\theta}_{i4}$ and $2 \pi k_{i}$ 
of the AP-SU(2) field profile (see, Eqs.~\eqref{eq:electric_field} 
and ~\eqref{eq:monopole_current}).
The DAH field profiles are calculated
with the same spatial volume as 
used in the SU(2) simulation, namely $32^{3}$,
imposing the same periodic boundary conditions
for all three directions.
The length of the DAH flux tube is taken 
equal to that of the AP-SU(2) flux tube to be fitted.
We extract the profile all over the midplane 
cutting the flux tube between the quark and the antiquark.
The DAH lattice spacing is assumed to be
the same as $a(\beta)$ of the SU(2) lattice.
Once the fit has found the optimal set of dimensionless
DAH  parameters, the physical masses are fixed 
with the help of $a(\beta)$.
To seek the set of parameters which provides
minimum $\chi^{2}$,
we use the MINUIT code from the CERNLIB.

\par
After getting the set of the DAH parameter,
we check whether this set can reproduce
the composed internal structure of the electric
field profile as a superposition of the Coulombic 
plus the solenoidal field by applying the Hodge decomposition 
as in Eq.~\eqref{eq:first} to the dual field strength.
Each field strength can be constructed from
the DAH photon link ($B^{ph} = 2 \pi \Delta^{-1} \delta \Sigma$)
and the DAH monopole link ($B^{mo}=\Delta^{-1} \delta F$),
respectively.
Note that the field strength from the photon (monopole) links
describe the Coulombic (solenoidal) electric field.

\subsection{Fitting results}

\par
We fit the flux-tube profiles 
at $\beta=2.5115$ from $W(3,6)$, $W(4,6)$,
$W(5,6)$,  and $W(6,6)$, and 
at $\beta=2.6$ from $W(4,8)$, $W(5,8)$, $W(6,8)$, 
$W(7,8)$, and $W(8,8)$.
Here, the physical length of the temporal extension
of the Wilson loop for these $\beta$ values 
is approximately the same, $0.48$ fm.
We did not attempt to 
fit the profiles from the Wilson loops with $R >T$,
since they still contain the contribution 
from excited states ($T$ dependence) even after the smearing
(see, Sec.~\ref{subsec:smearing} 
and Appendix~\ref{sec:smearing-detail}).
In the fit, we have taken into account the data
from $\rho/a \geq 2$ for $\beta=2.5115$
and  from $\rho/a \geq 3$ for $\beta=2.6$
to certain maximum radii which
provide the positive expectation values for
the field profiles.
We have checked that the DAH parameters
emerging from the fit
are rather insensitive with respect to restricting the fit range 
(further increasing the minimal radius).

\par
\begin{figure}[t]
\centering
\includegraphics[width=14cm]{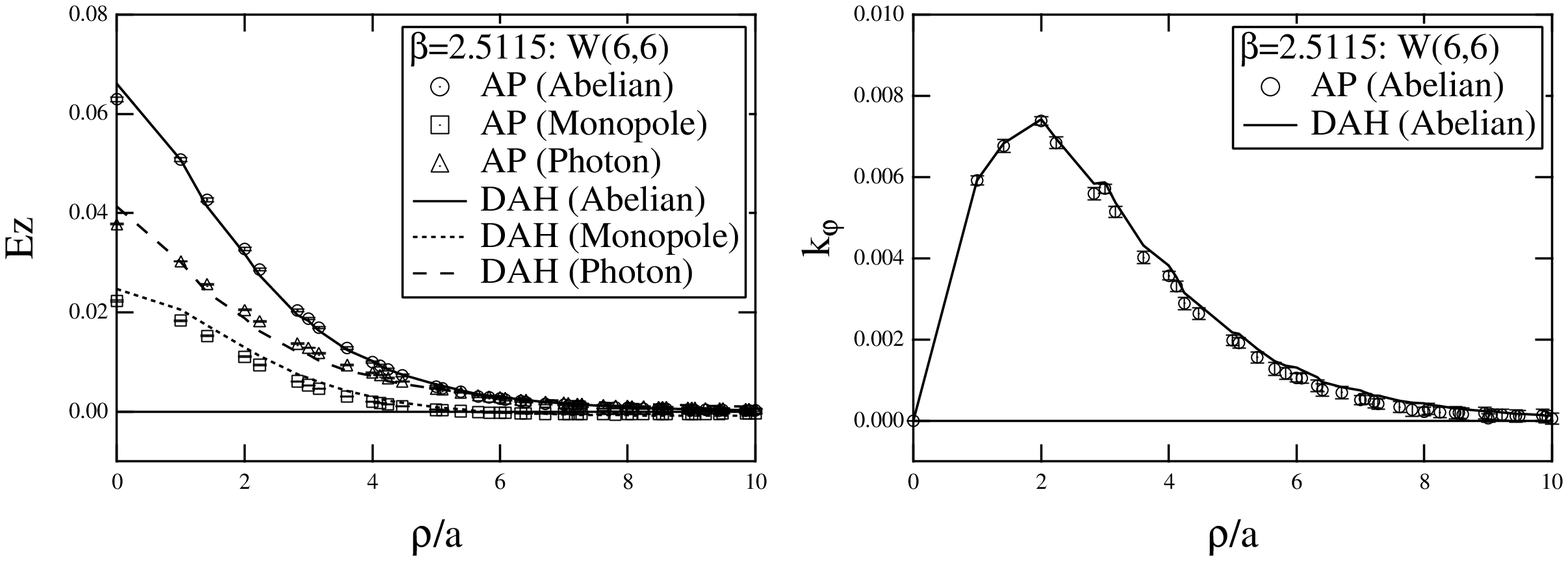}\\
\includegraphics[width=14cm]{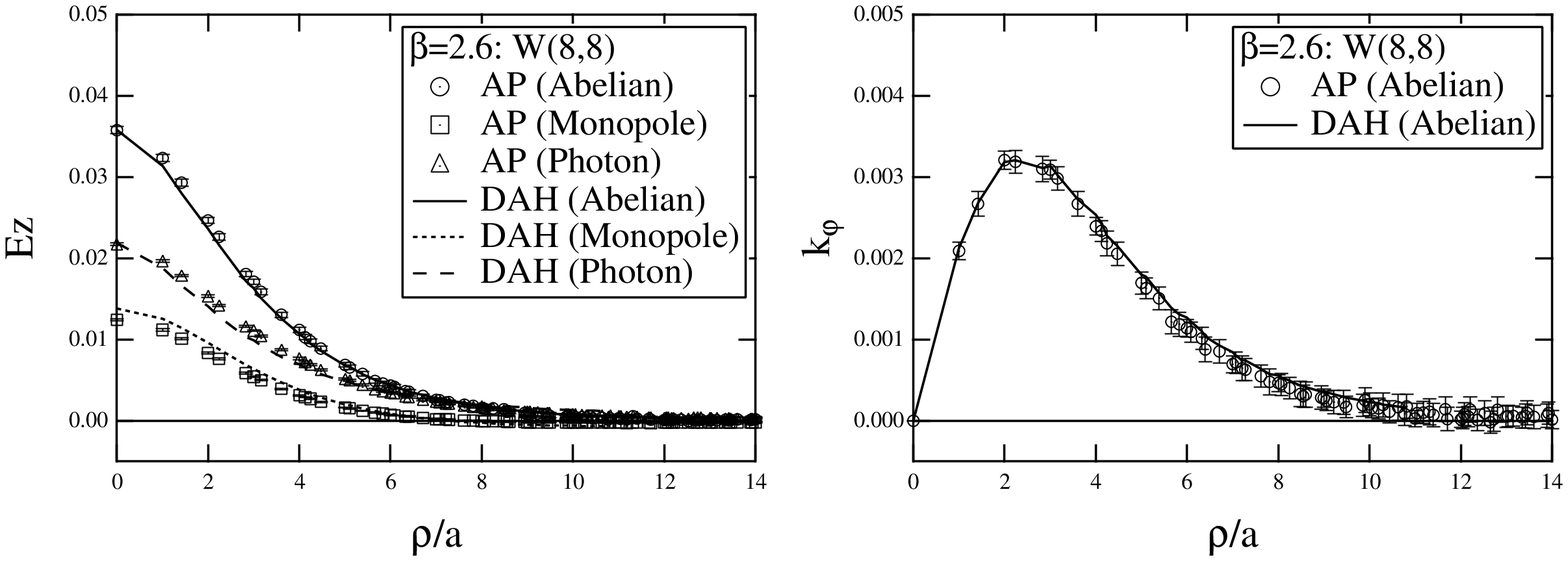}\\
\caption{Some examples of the fitting at 
$\beta=2.5115$ for $R=6$ (upper row), 
and $\beta=2.6$ for $R=8$ (lower row).
The solid line is the DAH flux-tube profile (obtained by the fit).
The dotted and dashed lines correspond to
its monopole and photon parts 
(as predicted using the fit parameters).}
\label{fig:fitsample}
\end{figure}
\begin{figure}[t]
\centering
\includegraphics[width=7.5cm]{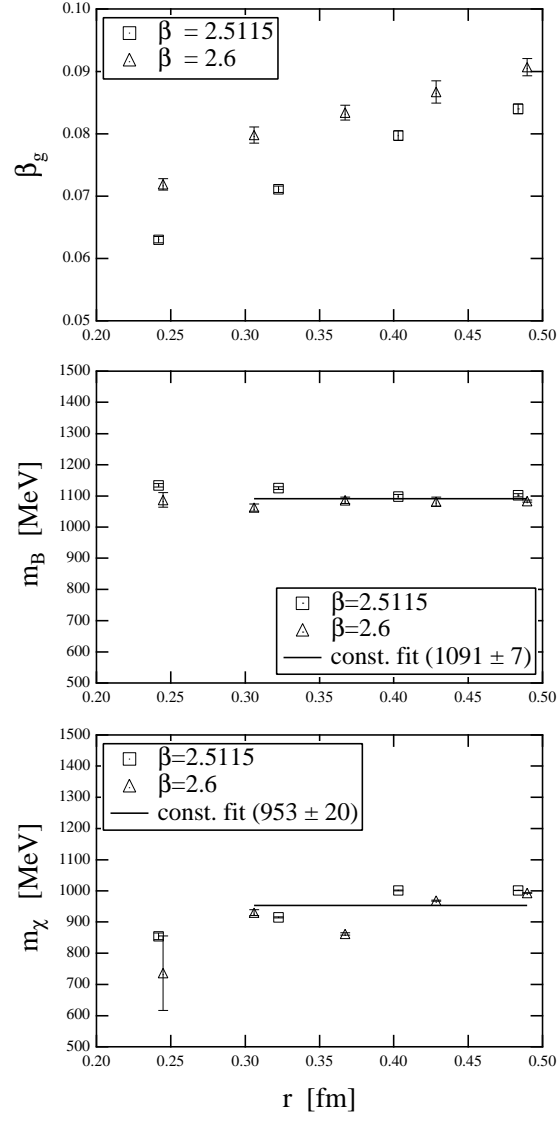}
\caption{The DAH parameters $\beta_{g}$, $m_{B}$ and $m_{\chi}$ as
functions of the physical $q$-$\bar{q}$ distance.}
\label{fig:fitparam_scale}
\end{figure}

In Table~\ref{tab:fitparam_org}, we summarize the 
parameters obtained by the fit.
In Fig.~\ref{fig:fitsample}, 
we show how the 
AP flux tube is described by the DAH one
using the profiles from
$W(6,6)$ at $\beta=2.5115$ and
from $W(8,8)$ at $\beta=2.6$.
One can see that the profiles 
from the Abelian Wilson loop are reproduced.
Remarkably, the resulting DAH parameters 
also reproduce the composed internal structure of the
AP flux tube as well.
In this sense, the fit which takes into account the
finite $q$-$\bar{q}$ distance works very well.
In Fig.~\ref{fig:fitparam_scale},
we plot the fitting parameters
as a function of the physical $q$-$\bar{q}$
distance, where the scale of the masses 
is recovered by using the SU(2) lattice spacing $a(\beta)$.
The maximum physical $q$-$\bar{q}$ distance is 
around 0.5 fm.
We find that the $\beta_{g}$ becomes large
as increasing  the $q$-$\bar{q}$ distance, $r = R a(\beta)$, 
while the masses of the dual gauge boson
and of the monopole are rather stable.
The constant fit of the masses using the stable data
for $r> 0.3$ fm provides
\bea
&&m_{B} = 1091 (7)  \; \mbox{MeV},\\
&&m_{\chi} = 953  (20)  \; \mbox{MeV}.
\eea
The GL parameter is then found to be
\bea
\kappa  = \frac{m_{\chi}}{m_{B}} = 0.87(2)  <1,
\eea
which means that the vacuum corresponds to weakly type I
in terms of the classification of the dual superconductivity.
However, we have noticed that the change of the dual gauge
coupling $\beta_{g}$ as a function of $r$ indicates
that the vacuum cannot completely be regarded as
the classical one.
In fact, if one defines 
an effective Abelian electric charge based on the 
Dirac quantization condition as
$e_{\rm eff}=4\pi/g=4 \pi \sqrt{\beta_{g}}$,
this coupling shows an anti-screening behavior;
$e_{\rm eff}$ becomes large 
with increasing $r$.
The constant behaviors of $m_{B}$ and $m_{\chi}$
indicate that various widths of the AP flux tube,
which are defined by the inverse of 
these masses (the penetration depth $m_{B}^{-1}$ and
the coherence length $m_{\chi}^{-1})$,
do not depend on $r$. 
This is established at least up to a $q$-$\bar{q}$ 
distance $r=0.5$ fm.

\noindent
\begin{table}[t]
      \centering
\caption{The $q$-$\bar{q}$ distance dependence 
of the DAH parameter}
\begin{tabular}{|l||l|l|l|l|l|l|l|}
    \hline
    $\beta$ &  $R$  &  $\beta_{g}$ & $m_{B}$ & $m_{\chi} $ & 
    $\chi^{2}$/dof       & fit range \\
     \hline \hline
  2.5115 & 3 & 0.0630(5)  & 0.4633(23) & 0.3490(43) & 135/67  &  
  $\rho/a \geq 2$ \\
        \hline	  
  2.5115 & 4 & 0.0711(5)  & 0.4595(17) & 0.3738(7)  & 99.9/81 &  
  $\rho/a \geq 2$ \\
        \hline
  2.5115 & 5 & 0.0797(8)  & 0.4485(29) & 0.4090(6)  &  77.4/81 & 
  $\rho/a \geq 2$ \\
        \hline
  2.5115 & 6 & 0.0840(8)  & 0.4504(21) & 0.4091(3)  & 186/97 & 
  $\rho/a \geq 2$ \\
        \hline\hline
  2.6 &      4 & 0.0719(9)  &  0.3372(72) & 0.2284(371) &  35.7/91 &
  $\rho/a \geq 3$ \\
        \hline
  2.6 &      5 & 0.0798(13) & 0.3295(36) & 0.2884(29)  &  22.0/91 &  
  $\rho/a \geq 3$ \\
        \hline
 2.6 &       6 & 0.0834(12) & 0.3368(31) & 0.2673(11) &  33.3/75 &  
 $\rho/a \geq  3$ \\
        \hline
 2.6 &       7 & 0.0867(18) & 0.3354(46) & 0.3004(5) &  37.3/91 &  
 $\rho/a \geq  3$ \\
        \hline
 2.6 &      8 &  0.0907(14) & 0.3360(10) & 0.3081(2) &  75.5/105 &  
 $\rho/a \geq 3$ \\
        \hline
\end{tabular}
\label{tab:fitparam_org}
\end{table}

\clearpage
\section{Summary and conclusions}
\label{sec:conclusion}

The main aim of this paper 
has been to present the flux-tube profile data within 
Abelian-projected (AP) SU(2) lattice gauge theory
in the maximally Abelian gauge (MAG)
in a quality and sufficiently detailed 
in order to warrant quantitative discussions
from the point of view of the dual Abelian Higgs (DAH) model.
We have mainly studied {\it i}) the lattice Gribov copy effect 
associated with the MAG fixing procedure
and {\it ii}) the scaling property ($\beta$ independence)
of the flux-tube profile 
using a large lattice volume, $32^{4}$.
During these investigations, we have always paid 
special attention to the composed internal structure of the AP flux tube.
We have also carefully monitored the effect of smearing.

\par
({\it i}) We have found that the flux-tube profile is
very sensitive to lattice Gribov copy problem in the MAG,
in particular, the monopole-related parts of the profile are 
strongly affected. 
The monopole current profile is overestimated
if one uses a non-improved gauge fixing algorithm 
or just one single gauge copy.
Since we do not know the real global maximum of 
$R[U^{g}]$ (see, Eq.~\eqref{eq:gauge_functional}), 
we cannot insist that our result is the final one.
However, we have obtained significantly 
corrected profile data 
by virtue of the over-relaxed 
simulated annealing algorithm
converging within a moderate number of gauge copies.

\par
({\it ii}) 
We have confirmed the scaling property of the flux-tube profile,
which has been achieved by using the well-gauge-fixed
configurations and by taking into account the finite $q$-$\bar{q}$ 
distance effect properly.
In fact, the flux-tube profile is strongly 
depending on the size ($R$ and $T$) of 
the Abelian Wilson loops $W(R,T)$.
At finite $q$-$\bar{q}$ distance $R$, 
the photon part of the flux-tube profile is crucially 
contributing to the total Abelian electric field 
measured at any distance from the external charges.

\par
Finally, we have investigated the effective parameters
of the U(1) dual Abelian Higgs (DAH) model by fitting 
the AP flux tube with the DAH flux-tube solution.
We have adopted a new fitting strategy;
in order to obtain the DAH flux-tube solution,
we have defined the DAH model on the dual lattice
and have solved the field equations numerically
with the same size of the spatial lattice volume
($32^{3}$) and with the same periodic boundary condition 
as in the AP-SU(2) simulations.
In particular, we have fully taken into account
the finite $q$-$\bar{q}$ distance effect.
As a result, we also could reproduce  the composed
internal structure of the AP flux tube in terms of the DAH one.
We have found that the dual gauge coupling 
depends on the $q$-$\bar{q}$ distance, which becomes large 
with increasing $q$-$\bar{q}$ distance.
In this sense, the comparison between the AP flux tube
and the DAH flux tube considered at the classical level is not perfect.
On the other hand, we have found that the masses 
of the dual gauge boson 
and the monopole are almost constant as a function of 
the $q$-$\bar{q}$ distance up to 0.5 fm, which 
take values of 1100 MeV and 950 MeV, respectively.
The Ginzburg-Landau parameter takes 
a value slightly smaller than unity, indicating that the vacuum 
is classified as weakly type-I dual superconductor.

\par
We should mention that, while we have chosen the DeGrand-Toussaint
prescription for the electric flux and the monopole currents featuring
in the description of flux tube in the Abelian projected gauge theory,
this choice is not unique.
We have made our choice because of the
possibility to relate then the AP theory to the DGL 
description~\cite{Koma:2003gq}.
Haymaker et al.~\cite{Cheluvaraja:2002yj}, 
on the other hand, have proposed an alternative
definition to satisfy the the Maxwell equations 
even at finite lattice spacing.
It would also be interesting to study the differences of the measured 
flux-tube profile with their definition.

\par
Suggested by an effective bosonic string description~\cite{Luscher:1981iy},
it is expected that the width of the flux tube 
broadens with increasing $q$-$\bar{q}$ distance.
If such an effect exists in the effective DAH description, too,
we would have to see the change of the DAH mass parameters
as a function of the $q$-$\bar{q}$ distance.
Our results show that, at least until 0.5 fm,
the width of the flux tube is 
appearing as an almost stable vacuum property.
It could be argued that the bosonic-string-like 
features of the flux tube
might become manifest only for much more elongated strings.
In order to study the existence of 
string roughening, one might be forced to
study the profiles correlated with Wilson loops of much larger size.
For larger $R$, if one takes small $T$,
one can get the signal of the flux-tube profile with the help of 
smearing techniques.
However, one has to care that
such a profile does not immediately 
correspond to the physical profile at $T \to  \infty$.
In order to check if the flux tube becomes broader,
the profile should be investigated in a
$T$-independent regime.

\par
It would be worthwhile to extend
the strategy of this paper to the AP-SU(3) flux tube
in order to discuss the quantitative relation 
between SU(3) gluodynamics and the U(1)$\times$U(1) DAH model.

\subsection*{Acknowledgment}

We are grateful to M.I.~Polikarpov for valuable 
discussions and the collaboration 
leading to our previous paper~\cite{Koma:2003gq}.
We wish to thank V.~Bornyakov, H.~Ichie, G.~Bali, 
P.A.~Marchetti, V.~Zakharov,
R.W.~Haymaker and T.~Matsuki 
for their constructive discussions.
We acknowledge also the collaboration of
T.~Hirasawa in an early stage of the present study.
Y.~K. and M.~K. are grateful to P.~Weisz for the 
discussion on the string fluctuation.
M.K. is partially supported by Alexander von Humboldt 
foundation, Germany.
E.-M.~I. acknowledges gratefully the support by Monbu-Kagaku-sho
which allowed him to work at Research Center for Nuclear Physics
(RCNP), Osaka University, where this work has begun.
He expresses his personal thanks to H.~Toki for the hospitality.
E.-M.~I. is presently supported by DFG through the DFG-Forschergruppe
'Lattice Hadron Phenomenology' (FOR 465).
T.~S. is partially supported by JSPS Grant-in-Aid for Scientific 
Research on Priority Areas No.13135210 and (B) No.15340073.
The calculations were done on the Vector-Parallel Supercomputer 
NEC SX-5 at the RCNP, Osaka University, Japan.

\appendix
\section{Fixing the smearing parameters}
\label{sec:smearing-detail}

\begin{figure}[t]
\includegraphics[width=14cm]{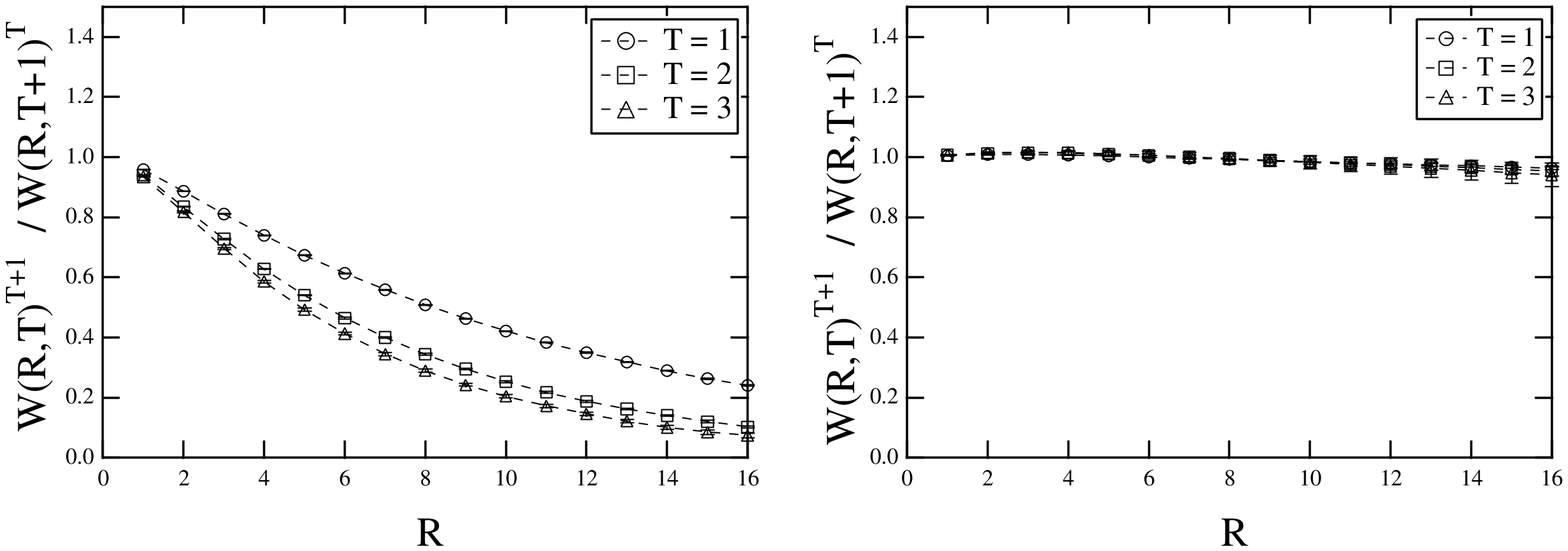}
\caption{The ground state overlap 
before (left) and  after (right) smearing at $\beta=2.5115$.}
\label{fig:c0_sme}
\end{figure}

\begin{figure}[t]
\includegraphics[width=14cm]{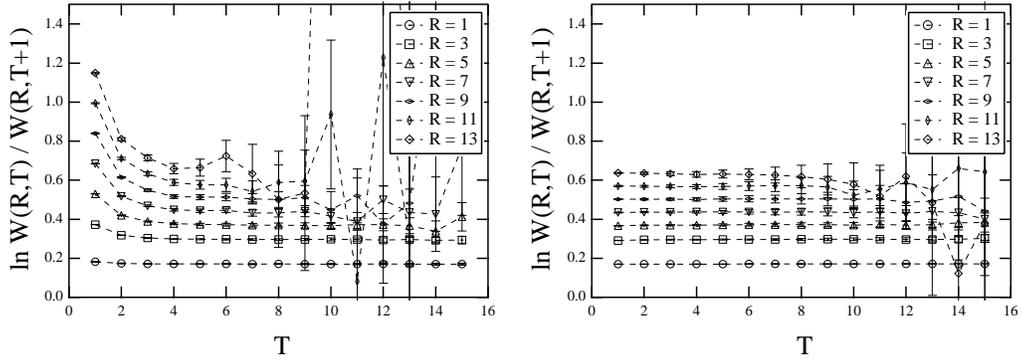}
\caption{The potential 
before (left) and after (right) smearing at $\beta=2.5115$.}
\label{fig:potential_sme}
\end{figure}

The smearing procedure for the spatial link variables as 
indicated in Eq.~\eqref{eq:smearing_step} successfully
reduces the contribution from excited states.
To find a set of optimized smearing parameters,
the weight $\alpha$ and the number of smearing sweeps
$N_{s}$, we need to investigate
the behavior of the ratio
\bea
C_{0} \equiv [W_{A}(R,T)]^{T+1}/[W_{A}(R,T+1)]^{T}
\label{eq:ground_state_overlap}
\eea
as a function of $R$ for some fixed values 
of $T$~\cite{Bali:1996dm}.
For interpretation, we notice that this ratio turns into the 
ground state overlap in the limit $T \to \infty$.
We apply the smearing step repeatedly on the spatial
links until we get a good ground state overlap.
In Fig.~\ref{fig:c0_sme} we show the typical behavior of 
the ground state overlap
{\it before} and {\it after} 
smearing at $\beta =2.5115$. We see clearly that $C_{0}$ got enhanced to
$C_{0} \approx 1$ 
as the result of smearing.
This justifies to select, for this $\beta$, 
the smearing parameters $\alpha=2.0$ and 
$N_{s}=8$ which have led to the improved ground 
state overlap.
Using this set, we can also confirm a 
typical improvement of the Abelian potential
\bea
V(R,T) = \ln (W_{A}(R,T)/W_{A}(R,T+1)) \; .
\label{eq:Abelian_potential}
\eea
This is shown in Fig.~\ref{fig:potential_sme},
where the potential is plotted as a function 
of $T$ for some fixed $R$.
After smearing, we see a clear pattern of plateaus 
ranging from $T=1$ to large $T$,
the height of which corresponds to the values of the
potential $V(R)$ at $T \to \infty$.

\par
\begin{figure}[t]
\includegraphics[width=14cm]{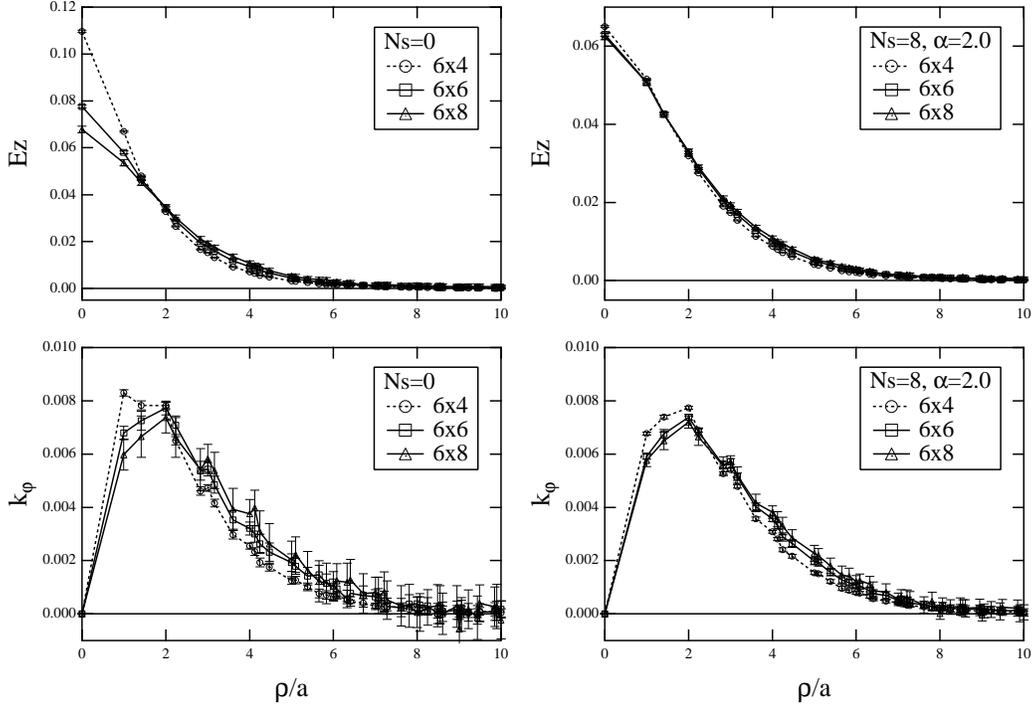}
\caption{
The flux-tube profiles before (left) and after (right) 
Abelian smearing at $\beta=2.5115$ for $R=6$ (fixed).
Three cases of the temporal length, $T=4$, 6 and 8, are shown.}
\label{fig:smearing_r6}
\end{figure}

In Fig.~\ref{fig:smearing_r6}, we show the effect 
of the smearing to the behavior of the flux-tube profile
at $\beta=2.5115$ for $W(6,4)$, $W(6,6)$, and $W(6,8)$.
Before smearing, the shape of the electric field and monopole 
current profiles  are depending on the size of the 
Wilson loop; the profile from the smaller Wilson loop 
is enhanced.
After the smearing, we see a reduction of the 
electric field for all cases and
at least, the profiles from $W(6,6)$, and $W(6,8)$
coincide within the numerical 
error (the error is also reduced).
This suggests that $T$-independence of the flux-tube profile is 
now achieved.
The shape of the monopole current profile is insensitive to 
smearing. However, we find a remarkable reduction of the error.
Note that the profile from $W(6,4)$ still 
did not converge into the same profile, which
means that the chosen smearing parameter set is not adequate in this case.

\section{Field equations of the lattice DAH model}
\label{sec:fieldeqs}

We note the field equations of the lattice DAH model.
For the dual gauge field,
\bea
\frac{\partial S}{\partial B_{i}(m)}
= 
\beta_{g} X_{i}(m) =0,
\eea
where
\bea
X_{i}(m) 
&=&
F_{ij} (m)+F_{ji} (m-\hat{j}) + F_{ik}(m) + F_{ki} (m-\hat{k})
- K_{i}(m)\; .
\eea
The last term corresponds to the monopole current
\bea
K_{i}(m)
&=&
- m_B^2 
\biggl [
\Phi^R (m) 
\left ( 
 \Phi^R(m+\hat{i}) \sin B_{i}(m)+\Phi^I (m+\hat{i})\cos B_{i}(m)
\right )
\nonumber\\*
&&
-\Phi^I (m)
\left ( 
 \Phi^R (m+\hat{i} )\cos B_{i}(m)-\Phi^I (m+\hat{i}) \sin B_{i}(m)
\right )
\biggr ] \; .
\eea
For the monopole fields,
\bea
&&
\frac{\partial S}{\partial \Phi^{R}(m)}
= 
\beta_{g} m_{B}^{2} X_{i}^{R}(m) =0 \; ,\\*
&&
\frac{\partial S}{\partial \Phi^{I}(m)}
= 
\beta_{g} m_{B}^{2} X_{i}^{I}(m) =0\; ,
\eea
where
\bea
X^R (m) &=& 
6 \Phi^R (m) 
-
\sum_{i=1}^3
\biggl \{
\left ( 
   \Phi^R (m+\hat{i}) \cos B_{i} (m)
- \Phi^I (m+\hat{i}) \sin  B_{i}(m)
\right ) 
\nonumber\\*
&&
+
\left ( 
\Phi^R (m-\hat{i}) \cos  B_{i} (m-\hat{i})+ 
\Phi^I (m-\hat{i}) \sin  B_{i}(m-\hat{i})
\right ) 
\biggr \}\nonumber\\*
&&
+
\frac{1}{2} m_\chi^2 \Phi^R(m) 
\left ( \Phi^{R}(m)^{2} + \Phi^{I} (m)^{2}-1 \right ) \;, \\
&&\nonumber\\
X^I (m)
&=& 
6 \Phi^I(m) 
-
\sum_{i=1}^3
\biggl \{
\left ( 
   \Phi^R (m+\hat{i})\sin  B_{i} (m)
   + \Phi^I (m+\hat{i}) \cos  B_{i}(m)
\right ) 
\nonumber\\*
&&
+
\left ( 
   \Phi^R (m-\hat{i}) (-\sin  B_{i} (m-\hat{i}) ) 
+ \Phi^I (m-\hat{i})\cos  B_{i}(m-\hat{i})
\right ) 
\biggr \}
\nonumber\\*
&&
+ \frac{1}{2} m_\chi^2 \Phi^I (m)
\left ( \Phi^{R}(m)^{2} +\Phi^{I}(m)^{2}-1 \right )  \; .
\label{eqn:feq-u1-lat2}
\eea


\end{document}